\documentclass[prl,twocolumn,amsmath,amssymb,floatfix,superscriptaddress,nofootinbib]{revtex4-1}

\usepackage{graphicx}
\usepackage{scrextend}
\usepackage{amssymb}
\usepackage{amsmath}
\usepackage{bm}
\usepackage{color}
\usepackage{multirow}
\usepackage{cprotect}
\usepackage{aas_macros}

\DeclareMathAlphabet\mathbfcal{OMS}{cmsy}{b}{n}
\definecolor{darkgreen}{RGB}{50,150,0}
\definecolor{purple}{cmyk}{0.5,1.0,0,0}

\def\edth{\;\raise1.0pt\hbox{$'$}\hskip-6pt\partial}
\def\baredth{\;\overline{\raise1.0pt\hbox{$'$}\hskip-6pt
\partial}}

\def\be{\begin{equation}}
\def\ee{\end{equation}}
\def\ben{\begin{equation} \nonumber}
\def\een{\end{equation}}
\def\ban{\begin{eqnarray*}}
\def\ean{\end{eqnarray*}}
\def\ba{\begin{eqnarray}}
\def\ea{\end{eqnarray}}
\def\({\left(}
\def\){\right)}

\newcommand{\bs}{\boldsymbol}
\newcommand{\spin}{\bs{j}}
\newcommand{\Msun}{M_\odot}

\definecolor{ultramarine}{rgb}{0.07, 0.04, 0.56}
\definecolor{cadmiumgreen}{rgb}{0.0, 0.42, 0.24}
\definecolor{indigo(dye)}{rgb}{0.0, 0.25, 0.42}
\usepackage[linktocpage=true]{hyperref}
\hypersetup{
colorlinks=true,
citecolor=ultramarine,
linkcolor=cadmiumgreen,
urlcolor=indigo(dye),
pdfauthor={},
pdftitle={},
pdfsubject={}
}

\begin{document}

\title{Observational detection of correlation between galaxy spins and initial conditions}

\author{Pavel Motloch}
\affiliation{Canadian Institute for Theoretical Astrophysics, University of Toronto, M5S 3H8, ON, Canada}

\author{Hao-Ran~Yu}
\affiliation{Department of Astronomy, Xiamen University, Xiamen, Fujian 361005, China}
\affiliation{Canadian Institute for Theoretical Astrophysics, University of Toronto, M5S 3H8, ON, Canada}
\affiliation{Tsung-Dao Lee Institute, Shanghai Jiao Tong University, Shanghai, 200240, China}

\author{Ue-Li~Pen}
\affiliation{Canadian Institute for Theoretical Astrophysics, University of Toronto, M5S 3H8, ON, Canada}
\affiliation{Tsung-Dao Lee Institute, Shanghai Jiao Tong University, Shanghai, 200240, China}
\affiliation{Dunlap Institute for Astronomy and Astrophysics, University of Toronto, M5S 3H4, ON, Canada}
\affiliation{Canadian Institute for Advanced Research, CIFAR Program in Gravitation and Cosmology, Toronto, M5G 1Z8, ON, Canada}
\affiliation{Perimeter Institute for Theoretical Physics, Waterloo, N2L 2Y5, ON, Canada}

\author{Yuanbo Xie}
\affiliation{Department of Astronomy, Beijing Normal University, Beijing, 100875, China}
\affiliation{Canadian Institute for Theoretical Astrophysics, University of Toronto, M5S 3H8, ON, Canada}

\begin{abstract}
\noindent

Galaxy spins can be predicted from the initial conditions in the early
Universe through the tidal tensor twist.  In simulations, their directions are well preserved through
cosmic time,
consistent with expectations of angular momentum conservation. We
report a $\sim 3 \sigma$ detection of correlation between observed oriented
directions of galaxy angular momenta and their predictions based on the initial density field
reconstructed from the positions of SDSS galaxies. The detection is driven by a
group of spiral galaxies classified by the Galaxy Zoo as (anti-)clockwise, with a modest
improvement from adding galaxies from MaNGA and SAMI surveys. This is
the first such
detection of the oriented galaxy spin direction, which
opens a way to use measurements of galaxy spins to probe fundamental physics in the early
Universe.

\end{abstract}

\maketitle

\section{Introduction}
\label{sec:intro}

Galaxy surveys such as \cite{1983ApJS...52...89H,2001MNRAS.328.1039C, 2000AJ....120.1579Y}
have taught us a great deal about the rules that govern our Universe. So far, our
knowledge has been based on measurements of galaxy positions and velocities. However, galaxies
also have rotational degrees of freedom, that we can utilize to learn more about the
Universe.

In papers \cite{Yu:2018llx,Yu:2019bsd} we pointed out that measurements of
galaxy angular momenta can be used to measure the sum of the
neutrino masses and to probe primordial chirality violations. Beyond these applications,
it seems possible to use galaxy spins to constrain primordial non-Gaussianity and
gravitational waves, though detailed prospects for these measurements are currently
unknown and under investigation (see \cite{Schmidt:2015xka,Biagetti:2020lpx} for similar
efforts using intrinsic alignments). Measurements of galaxy spins can be
also utilized to better constrain the initial conditions in the observed parts of the
Universe \cite{Lee:2000br,Lee:1999ii}, improving the reconstruction algorithms based on
galaxy positions and velocities \cite{2013MNRAS.432..894J,2014ApJ...794...94W,
2017PhRvD..96l3502Z,Schmittfull:2017uhh,Hada:2018fde,Zhu:2019gzu}. Finally, understanding
the origin of galaxy rotation can inform models of galaxy formation and provide a useful
input for studies of the intrinsic alignment and its effect on the weak lensing surveys
\cite{Kirk:2015nma}.

\begin{figure*}
\center
\includegraphics[width = \textwidth, trim={0 400pt 0 0}, clip]{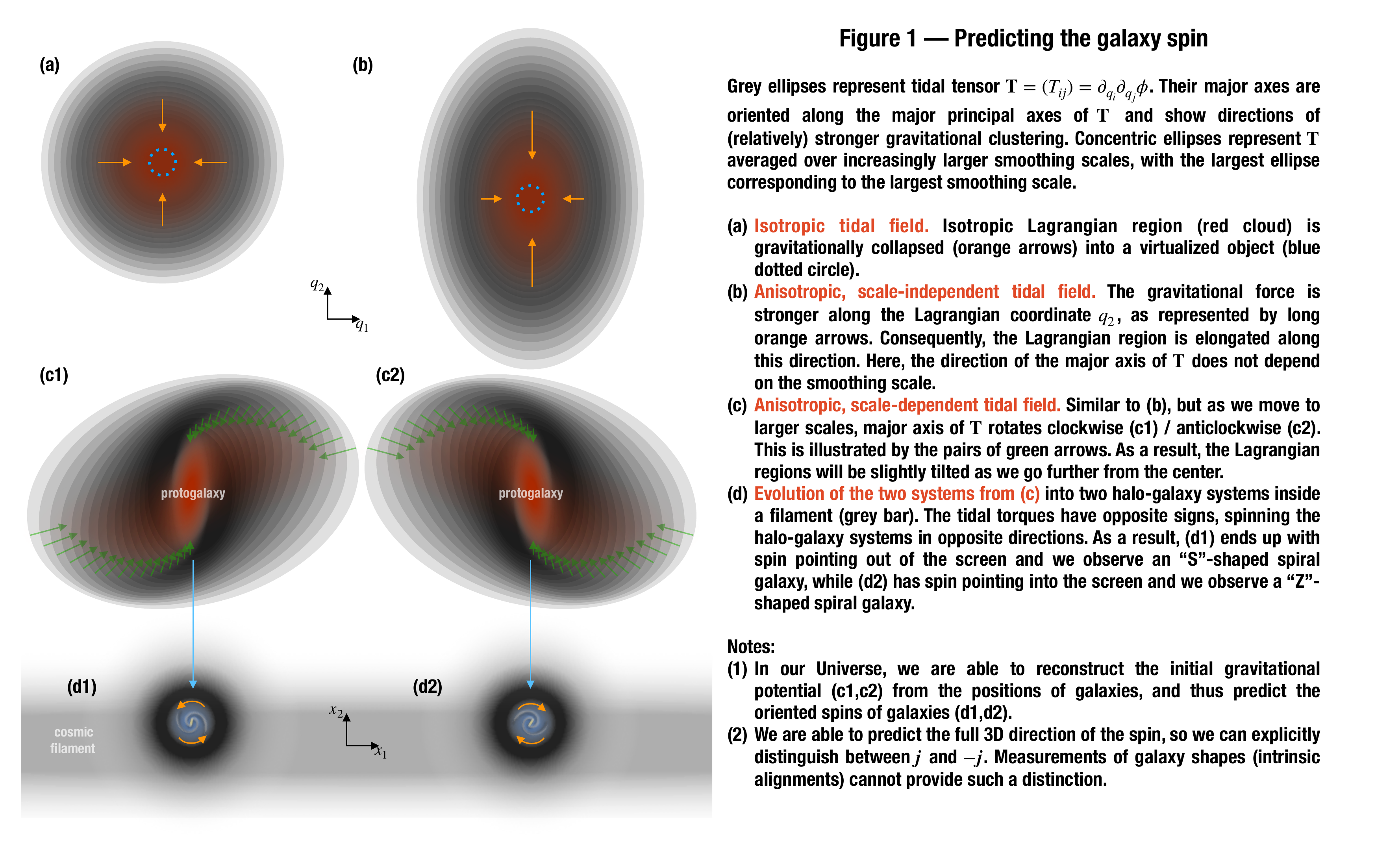}
\caption{Schematic illustration of how galaxy spins arise.}
\label{fig:protospin}
\end{figure*}

To be able to make good on any of the exciting prospects, it is necessary to connect the galaxy
spins with the conditions in the early Universe. 
In Fig.~\ref{fig:protospin} we present a schematic illustration of how galaxy spins
arise.
Tidal-Torque Theory \cite{1984ApJ...286...38W}
provides the basic mathematical framework,
relating the angular momentum of a dark matter (DM) halo to the initial misalignment of
the protohalo moment of inertia and the Hessian of the nearby gravitational
potential. In 
\cite{Yu:2019bsd} we found that \emph{directions}\footnote{Not necessarily the
amplitudes.} of the DM halo angular momenta\footnote{For brevity, we shorten ``direction
of angular momentum'' to \emph{spin}.} are well described by a second order quantity.
Because galaxy spins tend to align with the spins of their DM haloes
\cite{2015ApJ...812...29T, 2018arXiv180407306J}, it is expected that this second order
quantity can be used to predict galaxy spins as well.

In this paper, we study galaxy survey data to investigate to what
degree it is possible to observationally confirm the connection between galaxy spins and
initial conditions. To achieve this, we compare direct measurements of
galaxy angular momenta \emph{directions} with their predictions based on initial density
fluctuations reconstructed using a traditional reconstruction technique
\cite{2014ApJ...794...94W}.

Previous related studies (\cite{Kirk:2015nma} for a review,
\cite{Krolewski:2019bfv,Welker:2019puz}) have dealt either with intrinsic alignments of
galaxies or with alignments of galaxy spins and the large scale structure in their
neighborhood.
Unlike our work, that makes the connection to the initial conditions explicit, these earlier
studies only provide indirect evidence about the origin of galaxy spins. Additionally, we
consider the additional information contained in the orientation or the galaxy's spin, as
opposed to the earlier studies that focused on the \emph{unoriented} (headless)
directions.

\section{Summary of results from simulations}
\label{sec:sims}

In \cite{Yu:2019bsd} we
proposed that the \emph{direction} of
the angular momentum of DM halo $\spin_H$ can be well predicted from the
initial gravitational potential $\phi_i$ as
\be
\label{spin_formula}
	\spin_R=
	(j_{R\alpha})
	\propto
	\epsilon_{\alpha \beta \gamma}
	\(
	\partial_\beta \partial_\kappa
	\phi_i^R
	\)
	\(
	\partial_\kappa\partial_\gamma
	\phi_i^r
	\)
	,
\ee
where $\epsilon_{\alpha \beta \gamma}$ is the Levi-Civita symbol, $\phi_i^R$ (resp.
$\phi_i^r$) is the initial gravitational potential smoothed with a Gaussian kernel of
scale $R$ (resp. $r$) and repeated indices are summed over. The right hand side
of \eqref{spin_formula} is evaluated at the Lagrangian center of mass
of the halo. Geometrically, Eqn.~\eqref{spin_formula} measures the twisting of the
tidal tensor between scales $r$ and $R$, as shown in a 2-D projection in
Fig.~\ref{fig:protospin}.

Specifically, in N-body simulations we studied correlation strength, 
$
\mu_{RH} 
	\equiv 
	\Big\langle
		(\spin_R\cdot\spin_H)\big/|\spin_R||\spin_H|
	\Big\rangle
 ,
$
defined as mean cosine of the misalignment angle between the true spin of the halo
$\spin_H$ and the spin $\spin_R$ constructed according to \eqref{spin_formula}, averaged over 
haloes present in the simulation. 
$\mu_{RH}$ can range from $-1$ (vectors perfectly anti-aligned) to $1$ (perfectly
aligned) and for uncorrelated distributions $\mu_{RH} = 0$. For a random $\spin_R$,
$\mu_{RH}$ is uniformly distributed, with mean 0 and standard deviation $1/\sqrt{3}$.
For haloes heavier than $10^{12} \Msun$ we in our simulations found $\mu_{RH} \sim 0.5$,
which means that \eqref{spin_formula} predicts direction of the angular
momentum of a single DM halo with an uncertainty of $\sim 50^\circ$. 
This is roughly the same amount of information as if the halo spin was perfectly
predicted in 50\% of the cases and fully random in the other 50\% of the cases.
For comparison, weak
gravitational lensing typically has correlations at the percent level per object.

We found that optimal results are achieved with $R \rightarrow
r$. In this limit and using the Poisson equation, we can rewrite \eqref{spin_formula} as
\be
\label{limit_formula}
	(j_{R\alpha})
	\propto
	\epsilon_{\alpha \beta \gamma}
	\(
	\partial_\beta \partial_\kappa
	\rho_i^r
	\)
	\(
	\partial_\kappa\partial_\gamma
	\phi_i^r
	\) ,
\ee
where $\rho_i^r$ is the initial DM density smoothed with a Gaussian kernel of
scale $r$. This formula was earlier proposed as a proxy for galaxy spins in
\cite{Codis:2015tla}.  We can think of Eqn.~\eqref{limit_formula}
as estimating the instantaneous tidal twist at scale $r$.
In simulations we found that for perfectly known initial conditions $r$ should be near the
virial scale, i.e. $O(\mathrm{Mpc}/h)$, and mass dependent.
In practice, most galaxies surveys are sparsely sampled, limiting the
accessible values of $r$, with a resulting degradation in predicability.
We revisit the question of $r$ below.

\section{Data sets}
\label{sec:data}

\begin{figure}
\center
\includegraphics[width = 0.49 \textwidth, trim = {0 680pt 600pt 20pt}, clip]{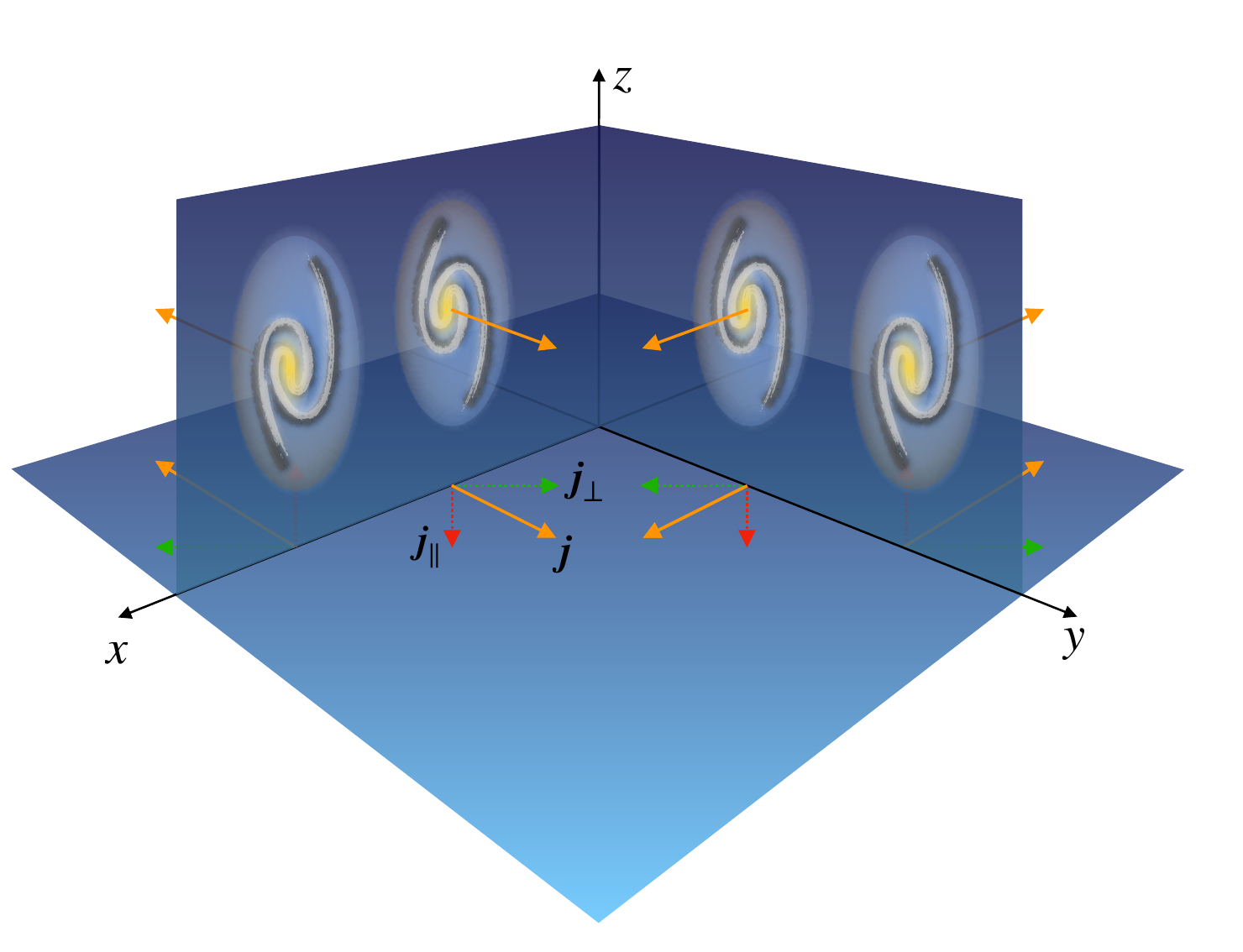}
\caption{
Components of galaxy spin.
Assuming spiral galaxies are thin disks with their spins perpendicular to the plane of the
disk, these four galaxies project onto the sky as identical ellipses despite all having
different 3D spin vectors (orange). For simplicity, we assumed all have their spin
oriented along either the $x$ or $y$ axis. The spins can be decomposed into components
parallel with ($\spin_\parallel$, red) and perpendicular to the line of sight
($\spin_\perp$, green). To break the four-fold degeneracy and determine the full 3D
orientation of spin, additional information is needed on top of the shape of the ellipse,
such as the sense of the spiral features (e.g. from the Galaxy Zoo) that can determine the
sign of $\spin_\parallel$, or the projected gas velocity along the line of sight (e.g.
form MaNGA, SAMI) that can determine $\spin_\perp$.
}
\label{fig:four_fold}
\end{figure}

In this section we summarize the data sets used in this work. We first describe the data set that
is used to calculate $\spin_R$, galaxy spins predicted from the initial conditions, and then
the data sets that are used to measure the galaxy spins $\spin_g$ directly. 
For most galaxies, we can currently only determine either the radial,
$\spin_{g\parallel}$, or perpendicular to the line of sight, $\spin_{g\perp}$, components
of their spin (see Fig.~\ref{fig:four_fold} for illustration).

\subsection{ELUCID}

We use the nearby Universe initial density field as determined by the ELUCID
collaboration. Here we only summarize their approach, details can be found in
\cite{2014ApJ...794...94W,Wang:2016qbz}.

They start with a catalog of galaxy groups \cite{Yang:2007yr} derived from the SDSS
\cite{2000AJ....120.1579Y} data
using a halo-based group finder. After determining masses of individual groups through
luminosity-based abundance matching, only groups in the Northern Galactic Cap, redshift
range $0.01 \le z \le 0.12$ and with masses above $10^{12}\Msun$ are retained. Peculiar
velocities are corrected to form a real space catalog. Using this group catalog, the space
is tessellated according to which galaxy group is the closest. Within such sub-volumes,
particles are placed randomly, in accordance with the expected density
profile of halo of given mass. This represents today's density field.

In the second step of the reconstruction, the best fit initial conditions are determined
using a Hamiltonian Monte Carlo method using a Particle-Mesh (PM) dynamics code. For
random initial conditions, the PM code is used to calculate the corresponding value of
today's density field. By comparing this density field
with that determined from the SDSS data, it is possible to construct a probability measure
on the space of the initial conditions. Due to inaccuracies of the PM code on small
scales, before comparison both density fields are smoothed on a scale of $4\, \mathrm{Mpc/h}$.
Iteratively probing the space of initial conditions then allows ELUCID to find the
initial conditions that best describe the local galaxy data.

From these best fit initial conditions, we use \eqref{limit_formula} to predict
the galaxy spins $\spin_R$. While we currently do not know what the optimal smoothing
scale $r$ to be applied to the data is and such study is beyond the scope of this paper,
the ELUCID pipeline provides us with a natural upper bound on the smoothing scale,
$r_\mathrm{max} = 4\, \mathrm{Mpc/h}$. From the
simulations we know that with known initial conditions, for
$10^{12}\Msun$ haloes that form the majority of our sample the best results are achieved with
$r \approx r_\mathrm{min} = 1\, \mathrm{Mpc/h}$. 
In what
follows, we will thus only consider smoothing scales between $r_\mathrm{min}$ and
$r_\mathrm{max}$ and search for $r$ that leads to the strongest correlation between
$\spin_R$ and the direct measurements of galaxy spins. We use a single, halo mass
independent, smoothing scale for the entire galaxy sample.

\subsection{Galaxy Zoo}

For spiral galaxies, sense of rotation of their spiral arms (clockwise or anti-clockwise,
i.e. in the sense of the letters Z or S) is closely related to the direction of the
angular momentum of the galaxy's gas. While about 4\% of galaxies have angular momentum
that is pointed in the opposite direction than that inferred from the orientation of the
spiral arms \cite{1982Ap&SS..86..215P}, this effect is not expected to bias the results. 
Determining the sense of rotation of a spiral galaxy is then a measurement of the sign of
the radial component of the galaxy's angular momentum (one bit of information), with
the galaxy's position vector aligned with its spin for the ``clockwise'' galaxies and
anti-aligned for the ``anti-clockwise'' galaxies.

We use data from Galaxy Zoo (GZ) \cite{Lintott:2008ne}, a citizen science project
where members of the public visually classified properties of almost $9 \times 10^5$
objects. ``Clockwise spiral galaxy'' and ``anticlockwise spiral
galaxy'' were among the six choices available, which gives us a catalog of galaxies with
a known sense of rotation. For each object, summary statistics of the voting results are
publicly available and we obtained them through CasJobs\footnote{https://skyserver.sdss.org/CasJobs/}.

For each object we label $F_\mathrm{(a)cw}$ the fraction of votes for the (anti-)clockwise
spiral. As our default sample, we choose those with $\max \(F_\mathrm{cw},
F_\mathrm{acw}\) \ge \mathcal{T}$, with $\mathcal{T} = 0.8$. We also investigate other
choices of $\mathcal{T}$ in what follows.

\subsection{MaNGA}

\begin{figure}
\center
\includegraphics[width = 0.49 \textwidth]{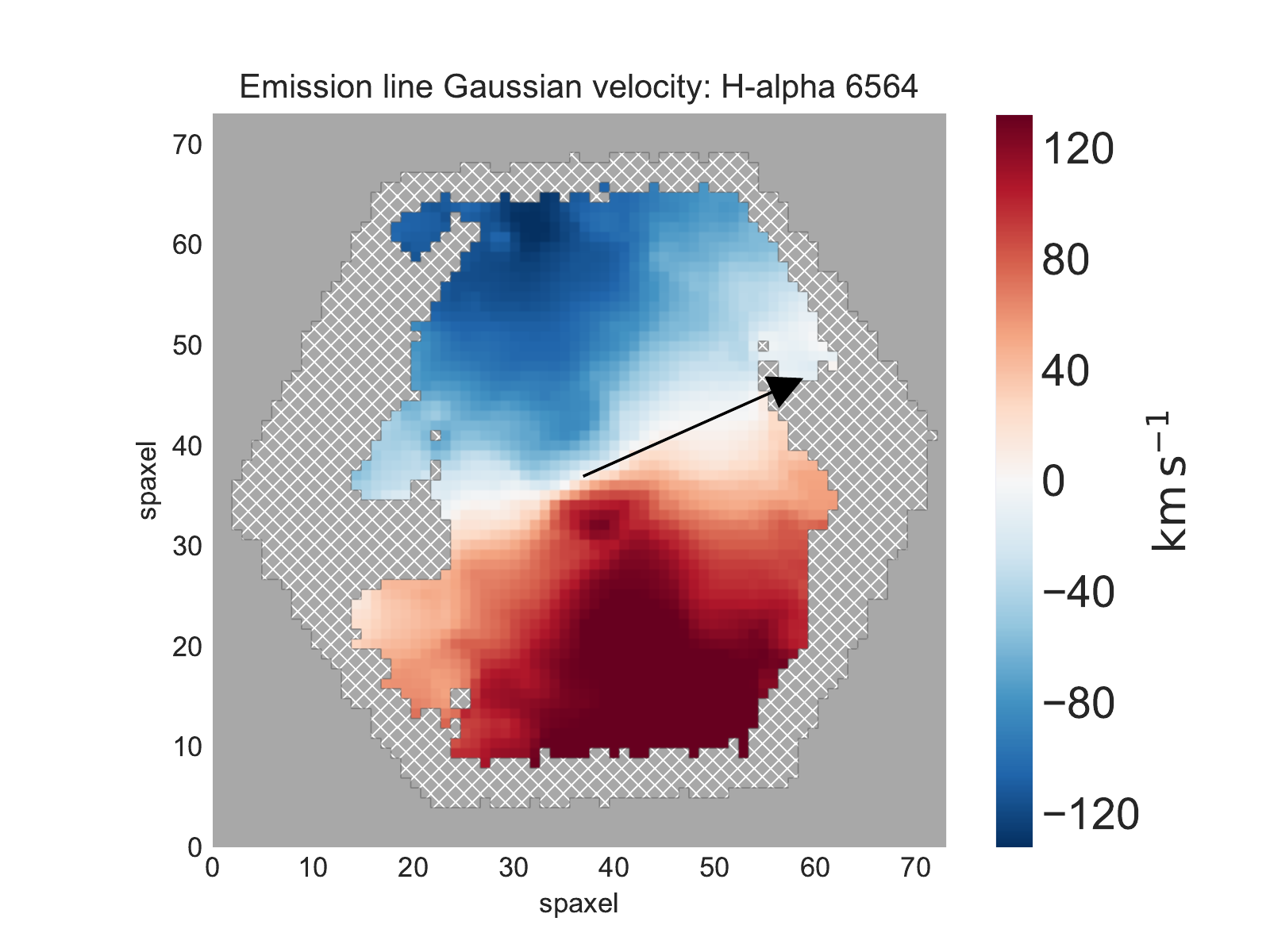}
\caption{Gas velocity, determined from the $H_\alpha$ line, for the galaxy SDSS
J024646.71-004343.7. The black arrow shows the (oriented) direction of the perpendicular
component of the galaxy's angular momentum $\spin_{g\perp}$ determined from the
gas velocity.} 
\label{fig:manga}
\end{figure}

Using the MaNGA survey \cite{Bundy:2014vpa}, it is possible
to determine orientation of $\spin_{g\perp}$ for several thousand galaxies.
For each galaxy, MaNGA measures integral field spectroscopy (IFS), which allows
determination of a 2D map of velocity of the baryonic
component (see Fig.~\ref{fig:manga} for an example). These maps can then be used to
determine the rotational state of individual galaxies. In this paper we use data that
is part of the SDSS DR15 \cite{Blanton:2017qot,Aguado:2018ynx}.

To analyze this data, we use Marvin\footnote{https://www.sdss.org/dr15/manga/marvin/}
\cite{2019AJ....158...74C} and focus exclusively on the $H_\alpha$ line.
For each galaxy we first construct mask from the measurements of $H_\alpha$ flux by
extending the default masks to further remove pixels with signal to noise ratio (SNR) below 10.
We then apply this extended mask on the 2D map of the $H_\alpha$-inferred gas velocity.
Because of the galaxy's rotation,
gas in approximately half of the galaxy is observed to be moving away from us, while the
gas in the opposite half of the galaxy is observed to be moving towards us, both relative
to the galaxy's overall motion.  The
projection of the axis of galaxy's rotation into the plane of the sky can be then
determined as the direction of zero velocity gradient in the map.

For each galaxy we assume the $H_\alpha$ gas velocity is well described by a linear
gradient in the angular coordinates and find the best fit direction of this gradient. The direction of
$\spin_{g\perp}$ is then set perpendicular to this
gradient, with the orientation determined by the regions of positive/negative velocity.
Using the known projection of $\spin_{g\perp}$ into the plane of the sky and the angular
position of the galaxy, we can determine the three-dimensional vector
$\spin_{g\perp}$ in galactic coordinates.
We neglect the uncertainty on the direction of $\spin_{g\perp}$, as it is
significantly smaller than the uncertainty on the reconstructed spin direction.

We perform several cuts. We first remove galaxies with less than 25 good pixels and those
that are blue shifted and are thus too close to be part of the volume probed by ELUCID. 
We then visually inspected gas velocity maps of the galaxies passing these preliminary
cuts and additionally remove 159 of them. These include galaxies for which it is difficult
to estimate their rotation, for example because of a more complicated morphology.

\subsection{SAMI}

SAMI Galaxy Survey \cite{2015MNRAS.447.2857B} is a collection of IFS data
obtained at the Anglo-Australian Telescope. We use
data\footnote{https://datacentral.org.au/services/download/}, specifically the ``Recommended-component
Ionised Gas Velocity'' maps, released in the DR2 of the survey \cite{2018MNRAS.481.2299S}.
Our analysis is analogous to that performed on the MaNGA data, only we do not mask pixels
with SNR $<$ 10. This presumably leads to a higher fraction of galaxies that are
removed upon visual inspection ($\sim 20\%$).

\subsection{Combining surveys and ELUCID cuts}

In the combined MaNGA and SAMI data, for 44 galaxies we have multiple observations. For
each such galaxy, we averaged $\spin_{g\perp}$ over all the observations. We checked 
these multiple measurements are mutually consistent.

With $\mathcal{T} =  0.8$, for about 600 galaxies we know both $\spin_{g\perp}$ and the
sign of $\spin_{g\parallel}$. If we assume these galaxies are very
thin disks, from known axis ratio it is in principle possible to construct full 3D vectors
$\spin_g$.  To avoid complications and given the relatively
small number of affected galaxies, for these galaxies we treat $\spin_{g\parallel}$ and
$\spin_{g\perp}$ independently, as if they described two different
galaxies. For each galaxy we then set either $\spin_{g\parallel}$ or $\spin_{g\perp}$ to
zero.

Looking only at galaxies in the volume for which ELUCID provides initial conditions and
those for which the closest galaxy group in the ELUCID catalog weights at least
$10^{12}\Msun$, we are left with 1609 galaxies for which we have measurement of
$\spin_{g\perp}$ and, for $\mathcal{T} = 0.8$ ($0.95$), with 13546 (5581) galaxies for
which we know $\spin_{g\parallel}$.

\section{Results}
\label{sec:results}

For the 15155 galaxies ($\mathcal{T} = 0.8$) for which we have direct
measurement of either $\spin_{g\parallel}$ or $\spin_{g\perp}$ we calculate the
correlation strength between the direct measurement of galaxy spin $\spin_g$ and its
prediction from the ELUCID initial conditions $\spin_R$ as
$
\cos \theta_\mathrm{Rg}
	\equiv 
	\(\spin_R\cdot\spin_g\)/|\spin_R||\spin_g| ,
$
where we use the displacement field determined by ELUCID to move $\spin_g$ to the Lagrangian
space.

\begin{figure}
\center
\includegraphics[width = 0.49 \textwidth]{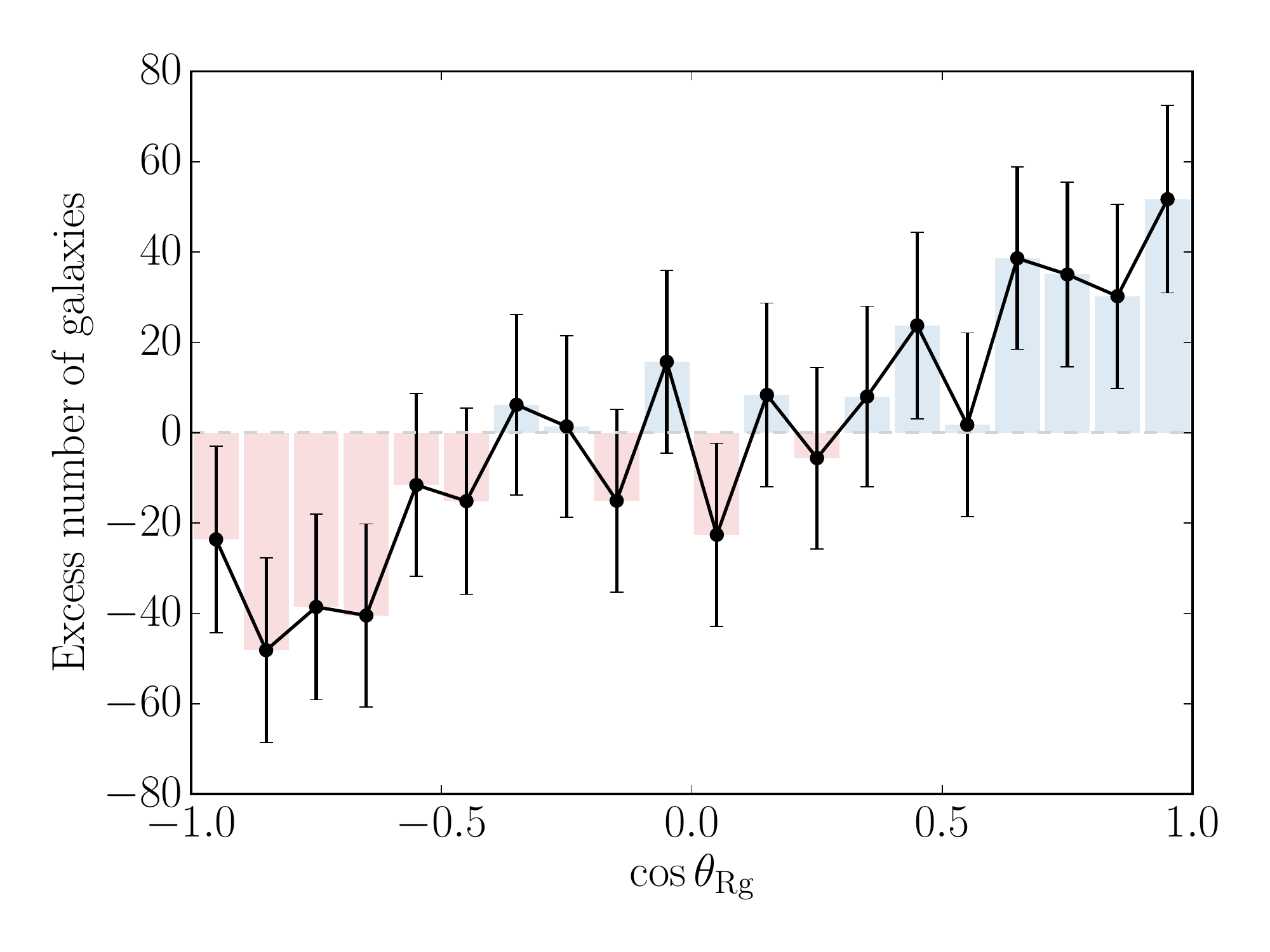}
\caption{
Excess of galaxies in bins of $\cos \theta_{Rg}$ over what would be expected if there was
no correlation between $\spin_g$ and $\spin_R$, the galaxy spins measured and predicted.
For our fiducial sample of 15155 galaxies and $r = 3\,\mathrm{Mpc/h}$. The (correlated)
error bars assume $\spin_g$ and $\spin_R$ are independent.
}
\label{fig:galaxy_excess}
\end{figure}

In Fig.~\ref{fig:galaxy_excess} we show excess number of galaxies in bins of $\cos
\theta_\mathrm{Rg}$ over what would be expected if there was no relationship between
$\spin_R$ and $\spin_g$.
This is for our fiducial sample and for the smoothing scale that leads to the strongest correlation between the
two sets of vectors, $r = 3\, \mathrm{Mpc/h}$.
We see a clear excess near  $\cos \theta_\mathrm{Rg} \approx
1$, where the two vectors are aligned, and deficit on the opposite end. The mean value averaged over all the
galaxies in our sample is $\mu_{Rg} \equiv \langle \cos \theta_\mathrm{Rg}\rangle =
0.017$.

\begin{figure}
\center
\includegraphics[width = 0.49 \textwidth]{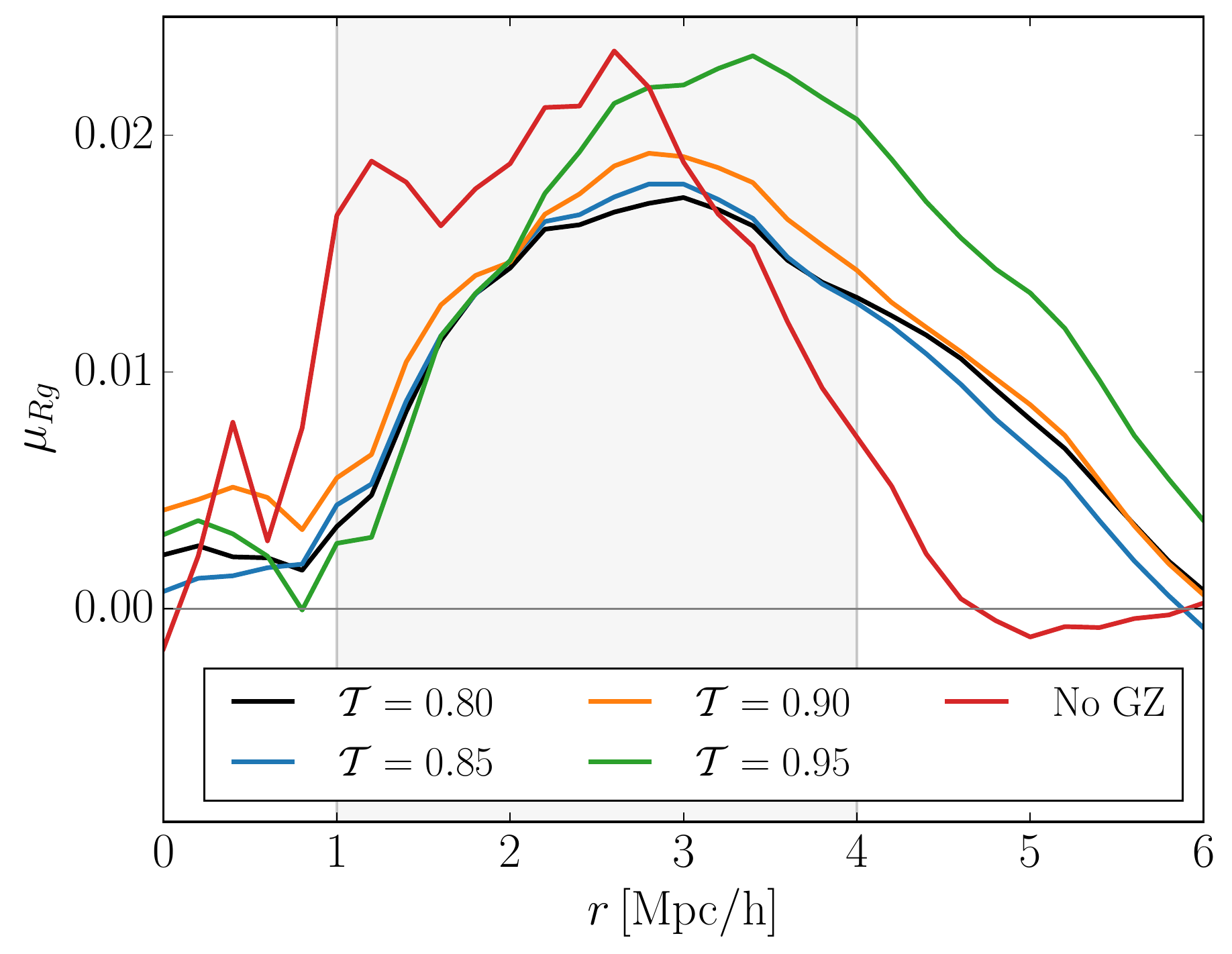}
\caption{Dependence of the detected correlation strength $\mu_{Rg}$ on the smoothing scale
$r$. Shown results for our default sample ($\mathcal{T} = 0.8$, black), together with
samples more restrictive on the GZ classification. Also shown results for only
the Manga and SAMI galaxies (red). The range of $r$ probed
when evaluating significance of the detection shown in gray.}
\label{fig:mu_on_r}
\end{figure}

In Fig.~\ref{fig:mu_on_r} we show dependence of $\mu_{Rg}$ on the smoothing scale for our
fiducial sample, together with several sub-samples where we are more strict on the GZ
classification. In
all samples we find that $r \sim 3.0\, \mathrm{Mpc/h}$ is close to optimal, with correlation
strengths between 0.015 and 0.025. We see a trend where being more restrictive on the
GZ sample leads to higher correlations. As expected, for small and large $r$ the correlation drops.

To evaluate probability we have observed just a random fluctuation, we simulate 40 000
random sets of galaxy spins and use them instead of $\spin_g$. To parallel what we do with
the data as closely as possible, for the galaxies from MaNGA and SAMI we only simulate
$\spin_{g\perp}$. For the galaxies from
GZ, we randomly shuffle the ``clockwise'' or ``anticlockwise'' classifications and
the $\spin_{g\parallel}$ component accordingly. This takes into account the bias of human
observers \cite{Land:2008vh}, and keeps the number of ``clockwise'' galaxies constant. For each
random set of galaxy spins we then find $r$ between $1\, \mathrm{Mpc/h}$ and $4\,
\mathrm{Mpc/h}$ that leads to the maximal correlation. This gives us a distribution of
$\mu_\mathrm{Rg}$ when there is no relationship between $\spin_R$
and $\spin_g$. This distribution is well described by a Gaussian, with mean
$\overline{\mu_{Rg}}
= 3.7 \times 10^{-3}$ and standard deviation $\sigma_{\mu_{Rg}} = 4.0 \times 10^{-3}$.
Notice the positive mean, caused by choosing the optimal $r$ in each case.
Possibility that the $\mu_{Rg}$ we see in the data is caused by chance
is thus ruled out at $3.4\sigma$.

\begin{figure}
\center
\includegraphics[width = 0.49 \textwidth]{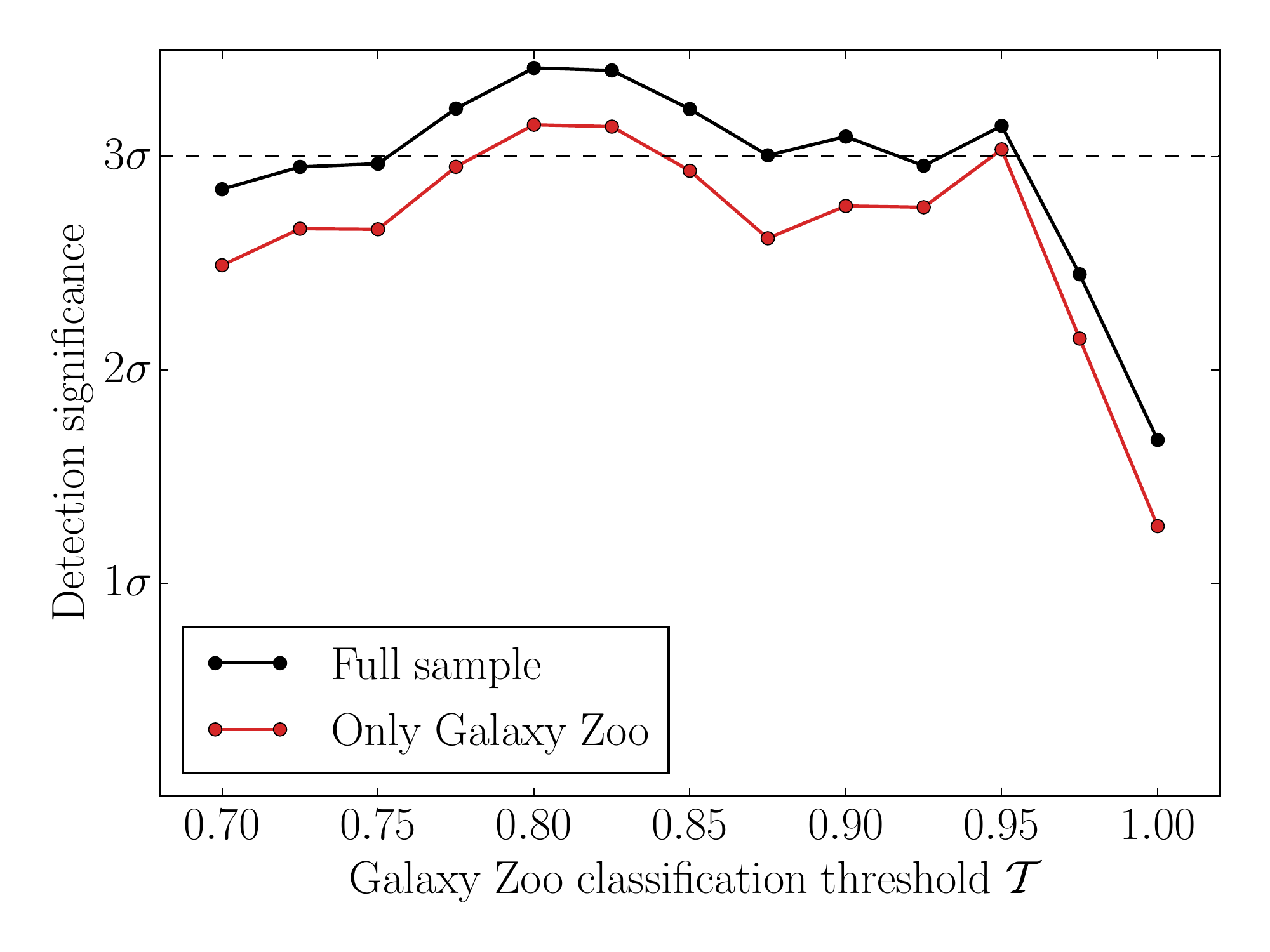}
\caption{Significance of the detection, as a function of the GZ classification
threshold $\mathcal{T}$. In red we show results without the MaNGA and SAMI galaxies.}
\label{fig:significance}
\end{figure}

In Fig.~\ref{fig:significance} we show dependence of the detection significance as a
function of the purity of the GZ sample $\mathcal{T}$. The significance is fairly
constant all the way to $\mathcal{T} = 0.95$,
after which it starts dropping, mainly due to the decreasing number of
galaxies and correspondingly increasing $\overline{\mu_{Rg}}$ and $\sigma_{\mu_{Rg}}$. The
detection significance is about 3.0$\sigma$, with majority of it coming from the GZ
galaxies. MaNGA and SAMI together only allow detection at a 1.0$\sigma$ level due
to the limited number of galaxies. Shrinking (extending) the range over which we
vary $r$ by
$1\, \mathrm{Mpc/h}$ increases (decreases) the detection significance by $\sim 0.1\sigma$.

\section{Discussion}
\label{sec:discussion}

In this paper we confirmed, with $\sim 3 \sigma$ significance, that formula
\eqref{limit_formula} can be used to predict directions of galaxy angular momenta.
While the correlation strength $\mu_{Rg}$ is only $\sim 0.02$, this is most likely
a consequence of poorly reconstructed initial conditions at $\sim 1\,
\mathrm{Mpc/h}$, the scale the DM halo spins are the most sensitive to.

With better reconstruction algorithms and new data, it should be possible to significantly
improve on this initial result. Especially DESI \cite{Levi:2013gra}, with its high density of
spectroscopically detected galaxies, is expected to push this science forward.
While DESI is ramping up, MaNGA and HECTOR \cite{2016SPIE.9908E..1FB} are expected to at
least double the catalog of galaxies with IFS. Classification of
galaxies into clockwise and anticlockwise then seems to be well suited for application of
machine learning techniques, which could potentially also increase the galaxy sample and thus
the detection significance.

In principle, the signal could be caused by a systematic effect that correlates the
reconstructed density field with the direct measurements, especially the
clockwise/anticlockwise classification. While we do not know of any such systematic, one should
keep this possibility in mind. With larger data sets and automated classification, ruling
out this issue might become possible.

With this in mind, establishing connection between initial conditions and galaxy
spins allows us to start probing exciting physical phenomena in the early Universe.

\acknowledgements{
We thank Huiyuan Wang for providing the catalog of ELUCID galaxy groups and useful
discussions.
}

\appendix

\bibliography{galspins_obs}

\end{document}